\documentclass{desyproc}
\usepackage{floatrow}
\begin{document}
%------------------------------------
\title{Gravity Resonance Spectroscopy and Einstein-Cartan
  Gravity}

%for single authors the superscripts are optional
\author{{\slshape Hartmut Abele$^{1}$, Andrei Ivanov$^{1}$, Tobias Jenke$^{1}$, Mario Pitschmann$^{1}$, Peter Geltenbort$^2$}\\[1ex]
$^1$Atominstitut, Technische Universit{\"a}t Wien Stadionallee 2, 1140 WIEN, Austria\\
$^2$Institut Laue Langevin, 71 avenue des Martyrs, 38000 GRENOBLE, France}

% if the proceedings are available online (e.g. at Indico)
% please enter the contribution ID or file_name below for the DOI
%\contribID{32}
\contribID{familyname\_firstname}

% TO THE CONFERENCE EDITORS:
% please update the following information
% before sending the template to the authors
\confID{11832}  % if the conference is on Indico uncomment this line
\desyproc{DESY-PROC-2015-02}
\acronym{Patras 2015} % if you want the Acronym in the page footer uncomment this line
\doi  % if there is an online version we will register DOIs

\maketitle

\begin{abstract}
The \textit{q}\textsc{Bounce} experiment offers a new way of looking at gravitation based on quantum interference. An ultracold neutron is reflected in well-defined quantum states in the gravity potential of the Earth by a mirror, which allows to apply the concept of gravity resonance spectroscopy (GRS). This experiment with neutrons gives access to all gravity parameters as the dependences on distance, mass, curvature, energy-momentum as well as on torsion. Here, we concentrate on torsion.
\end{abstract}

\section{Introduction}
\begin{figure}
\floatbox[{\capbeside\thisfloatsetup{capbesideposition={right,bottom},capbesidewidth=6cm}}]{figure}[\FBwidth]
{\caption{Pico-eV energy eigenstates $E_1$ to $E_5$ and Airy-function
  solutions of the Schr\"odinger equation for bound ultra-cold
  neutrons in the linear gravity potential of the Earth. The energy
  eigenstates are used for gravity resonance spectroscopy and the
  observed transitions between energy eigenstates are indicated by
  black arrows.}\label{fig:test}}
{\includegraphics[height=0.3\textheight]{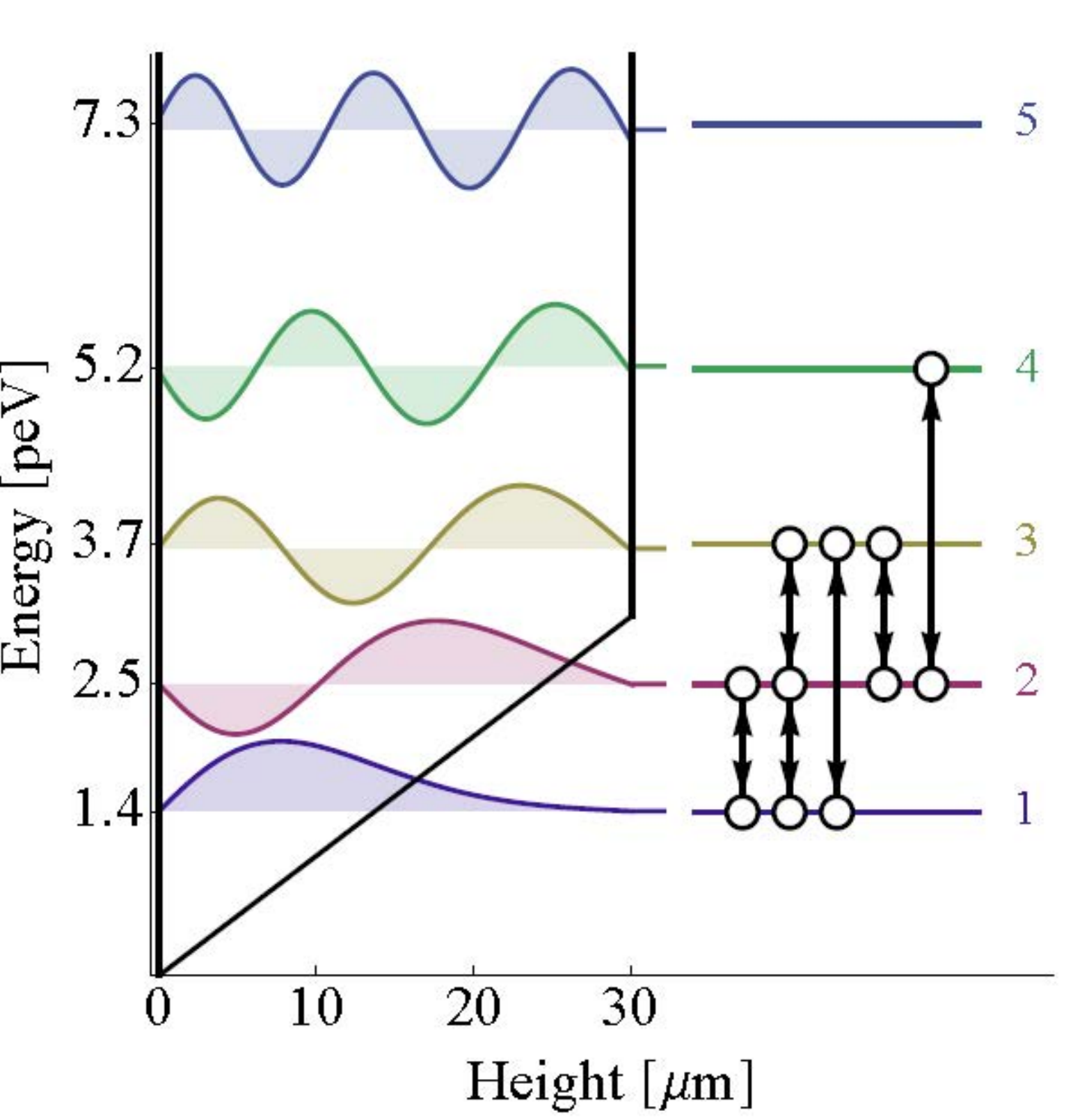}}
\label{fig:qboun}
\end{figure}

In the past few years, the \textit{q}\textsc{Bounce} collaboration has developed a new quantum-technique based on ultra-cold
neutrons. Due to their quantum nature, neutrons can  be manipulated in
novel ways for gravity research. For that purpose a gravitational resonance spectroscopy (GRS) technique has been implemented
 to measure the discrete energy eigenstates of ultra-cold neutrons in the gravity
potential of the Earth, see Fig.\,\ref{fig:qboun}. The energy levels
are probed, using neutrons bouncing off a horizontal mirror with
increasing accuracy. In 2011~\cite{Abele12}, we demonstrated that such
a resonance spectroscopy can be realized by a coupling to an external
resonator, i.e., a vibrating mirror.  In 2014, the first precision measurements of gravitational quantum
states with this method were presented. The energy differences between eigenstates shown in
Fig.\,\ref{fig:qboun} are probed with an energy resolution of
10$^{-14}$ eV. At this level of precision, we are able to provide
constraints on any possible gravity-like interaction. Then, we
determined experimental limits, first, for a prominent quintessence theory (chameleon fields) and, second, for axions at short
distances~\cite{Jenke2014}. Detailed information on an experimental realization of the quantum bouncing ball by measuring the neutron density distribution given by the wave function can be found in ~\cite{Abe09,Jen09}. The demonstration of the neutron's quantum states in the gravity potential of the Earth has been published in~\cite{Abele4,Abele10}.
%\begin{figure}
%\centering \includegraphics[height=0.30\textheight]{Wavefunction.pdf}
%\caption{Pico-eV energy eigenstates $E_1$ to $E_5$ and Airy-function
%  solutions of the Schr\"odinger equation for bound ultra-cold
%  neutrons in the linear gravity potential of the Earth. The energy
%  eigenstates are used for gravity resonance spectroscopy and the
%  observed transitions between energy eigenstates are indicated by
%  black arrows.}
%\label{fig:qboun}
%\end{figure}

It is planned to extend the sensitivity of this method to
an energy resolution  of 10$^{-17}$eV, and in the long run to
10$^{-21}$eV. The resonance spectroscopy method will therefore be extended to
a Ramsey-like spectroscopy technique~\cite{Abele2}.

At this level of sensitivity, the experiment addresses some important
problems of particle, nuclear and
astrophysics: Three of the most important current
theoretical and experimental problems of cosmology and particle
physics are i) the current phase (late--time) acceleration
of the expansion of the Universe
\cite{Perlmutter1997,Riess1998,Perlmutter1999}, ii) the nature of dark
energy, which accounts for about $69\,\%$ of the density in the Universe,
i.e. $\Omega_{\Lambda} \approx 0.69$ \cite{PDG2014,Copeland2006}, and
iii) the possible existence and nature of torsion, providing a basis for e.g.
Einstein--Cartan gravity \cite{Cartan1922,Cartan1923,Cartan1975,Kibble1961,Sciama1962}. One of
the simplest explanations for the acceleration of the expansion of the Universe
and dark energy is the introduction of the cosmological constant
\cite{Copeland2006}, which was introduced for the first time
in 1917 by Einstein in his paper ``{\it Cosmological Considerations in
the General Theory of Relativity}'' \cite{Einstein1917}. Einstein's original motivation,
outdated by Hubble's discovery of the expansion of the Universe soon afterwards,
was to obtain a static solution for the Universe.
However, modern quantum field theories naturally connect
the cosmological constant with the vacuum--energy of quantum fields.
To account for the experimentally observed expansion of the Universe consistent with
theories of the history of the Universe, the so-called chameleon scalar fields have been introduced.
To avoid any conflict with observations at terrestrial and solar system scales, the
properties of these new chameleon fields have to depend on the environmental density.
Especially, the effective mass of the chameleon field, and therefore the effective range of its interactions,  depend on
the density of the environment \cite{Chameleon1,Chameleon2}. The chameleon field is a specific
realization of {\it quintessence} \cite{Tsujikawa2013}. The chameleon field
as a source of dark energy has been discussed in \cite{Jain2013}.
%As a tool for such a test we propose to use the neutron
%interferometry and the gravity resonance spectroscopy with ultracold
%neutrons (UCNs) \cite{Abele2}--\cite{Jenke2014}.
\section{Einstein--Cartan Gravity}
In 1922 - 1925 Cartan proposed a theory \cite{Cartan1922,Cartan1923},
which is an important generalization of Einstein's general theory of
relativity \cite{Cartan1975}. In contrast to general relativity,
Einstein--Cartan theory allows space--time to have torsion in addition
to curvature, which may in principle couple to a particle spin. For a
long time Einstein--Cartan theory was unfamiliar to physicists and did
not attract any attention. In the beginning of the '60s of the last
century the theory of gravitation with torsion and spin was
rediscovered by Kibble \cite{Kibble1961} and Sciama
\cite{Sciama1962}. From the 1970s on, Einstein--Cartan theory has been
intensively investigated \cite{Hehl1995,Ivanov2015}. Recently,
it has been shown \cite{Ivanov2014} that in the non--relativistic
approximation of the Dirac equation in the effective gravitational
potential of the Earth, a torsion--matter interaction naturally
appears after taking into account also chameleon fields.  Such a
result demonstrates that chameleon fields can also serve as an origin
of space--time torsion. Gravity with torsion, caused by a scalar
field, was  discussed in detail by Hammond in
the review paper \cite{Hammond2002}.

In Einstein--Cartan gravity torsion appears as the antisymmetric part of
the affine connection \cite{Hehl1995}. Thus, torsion is an additional natural geometrical quantity
characterizing space--time geometry through spin--matter interactions \cite{Hehl1995}--\cite{Ivanov2015}.
It allows to probe the rotational degrees of freedom of space--time in terrestrial laboratories.
Torsion may be described by a third rank tensor ${\cal T}_{\alpha\mu\nu}$,
which is antisymmetric with respect to last two indices ${\cal
  T}_{\alpha\mu\nu} = - {\cal T}_{\alpha\nu\mu}$. It can be
represented in the following general form \cite{Kostelecky2008}:
${\cal T}_{\alpha\mu\nu} = \frac{1}{2}(g_{\alpha\mu}{\cal T}_{\nu} -
g_{\alpha\nu}{\cal T}_{\mu}) -
\frac{1}{6}\,\varepsilon_{\alpha\mu\nu\beta}{\cal A}^{\beta} + {\cal
  M}_{\alpha\mu\nu}$, where $g_{\alpha\sigma}$ and
$\varepsilon_{\alpha\mu\nu\beta}$ are the metric and the Levi--Civita
tensor, respectively. It possesses 24 independent degrees of freedom,
which are related to a 4--vector ${\cal T}_{\mu}$, a 4--axial--vector
${\cal A}_{\mu}$ and a 16--tensor ${\cal M}_{\alpha\mu\nu}$. The tensor degrees of freedom ${\cal
  M}_{\alpha\mu\nu}$ obey the constraints $ g^{\alpha\mu}{\cal
  M}_{\alpha\mu\nu} = \varepsilon^{\sigma\alpha\mu\nu}{\cal
  M}_{\alpha\mu\nu} = 0$.
A minimal inclusion of torsion in terms of the affine connection leads
to torsion--matter interactions, caused by the 4--axial degrees of
freedom only. As it has been shown in
\cite{Laemmerzahl1997,Kostelecky2008,Obukhov2014} the effects of
the torsion axial--vector degrees of freedom are extremely small. An
upper bound of $(10^{-22} - 10^{-18}  )\,{\rm eV}$
has been obtained from the null results on measurements of
Lorentz invariance violation.  Recent measurements of neutron spin
rotation in liquid ${^4}{\rm He}$, carried out by Lehnert {\it et al.}
\cite{Snow2014}, have lead to the upper bound $ |\zeta| < 5.4\times
10^{-5}\,{\rm eV}$ on a parity violating linear combination of the
time--components of the vector ${\cal T}_{\mu}$ and the axial--vector
${\cal A}_{\mu}$. Since the order of the
time--component of the torsion axial--vector is
about $10^{-18}\,{\rm eV}$ \cite{Kostelecky2008}, an enhancement of
the torsion--spin--neutron parity violating interaction can be
attributed to a contribution of the time--component of the torsion
vector ${\cal T}_{\mu}$. Unfortunately,
interactions of both the torsion vector ${\cal T}_{\mu}$ and the
torsion tensor ${\cal M}_{\alpha\mu\nu}$ degrees of freedom can be
introduced only phenomenologically in a non--minimal way
\cite{Kostelecky2008}. This diminishes a little bit the predicting power
of the experimental data \cite{Snow2014}, since the experimental
quantity $\zeta$ depends on some set of phenomenological parameters
multiplied by the time--components of the torsion vector ${\cal T}_0$,
and axial--vector, ${\cal A}_0$.  Nevertheless, the
experimental upper bound by Lehnert {\it et al.} \cite{Snow2014} can
be accepted as a hint on a possible dominance of the torsion vector
degrees of freedom ${\cal T}_{\mu}$ over the torsion
axial--vector ones ${\cal A}_{\mu}$.
\section{The \textit{q}\textsc{Bounce} Experiment}
Concerning chameleon fields, the corresponding solutions of the non-linear equations of
motion confined between two mirrors have been obtained
in~\cite{Ivanov2013} and used in~\cite{Jenke2014} in the
extraction of the contribution to the transition frequencies of
quantum gravitational states of ultra-cold neutrons (UCNs).

Furthermore, the development of a
version of Einstein--Cartan gravity with the torsion vector
${\cal T}_{\mu}$ degrees of freedom introduced in a minimal way
becomes meaningful and challenging. Clearly, such an extension of general relativity must not
contradict well--known data on the late--time acceleration of the
expansion of the Universe and dark energy dynamics. A possible route is using our results \cite{Ivanov2014}
and taking the torsion vector components ${\cal T}_{\mu}$ as the gradient of the
chameleon field. Such a version of a torsion gravity theory allows
to retain all properties of the chameleon field, which are necessary
for the explanation of the late--time acceleration of the Universe
expansion, dark energy dynamics and the equivalence principle
\cite{Will1993} (see also \cite{Chameleon1,Chameleon2}) and to extend
them by chameleon--photon and chameleon--electroweak boson
interactions, introduced in a minimal way.
\begin{figure}
\centering
\includegraphics[height=0.20\textheight]{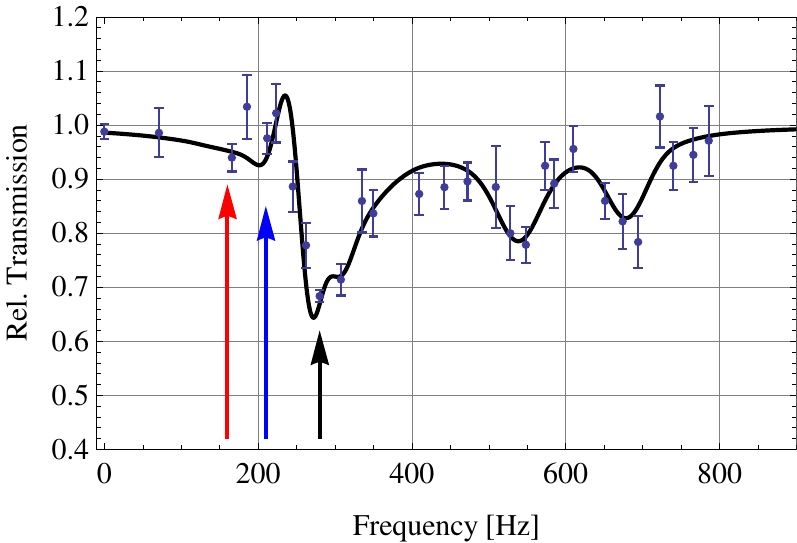}
\includegraphics[height=0.20\textheight]{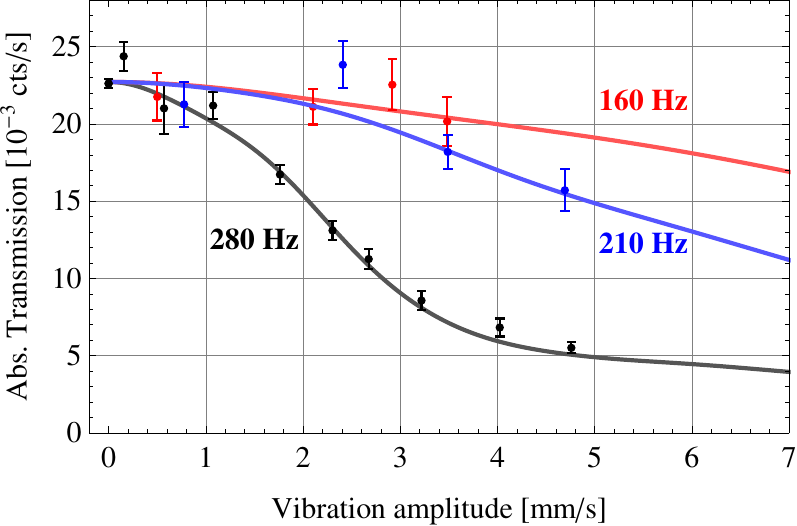}
\caption{Results for the employed GRS. Left: The transmission curve determined from the neutron count rate behind the mirrors as a function of oscillation frequency showing dips
corresponding to the transitions shown in Fig~\ref{fig:qboun}. Right: Upon resonance at 280 Hz, the transmission decreases with the oscillation
amplitude in contrast to the detuned 160 Hz. Because of the damping, no revival occurs. A detailed description of the
experiment can be found in~\cite{Jenke2014}.}
\label{fig:qbounce}
\end{figure}

For the experimental analysis of these chameleon induced torsion - matter
interactions very sensitive experiments are needed, which need
to overcome the barrier of extremely small magnitudes of the
torsion degrees of freedom.  As has been pointed out in
\cite{Brax2011,Ivanov2013} and proved experimentally in
\cite{Jenke2014}, UCNs, bouncing in the gravitational field of the
Earth above a mirror and between two mirrors can be a good laboratory
for testing chameleon--matter field interactions. The quantum
energy scale of UCNs is $\varepsilon = m g \ell_0 = 0.602\,{\rm peV}$,
where $m$, $g$ and $\ell_0$ are the neutron mass, the Newtonian
gravitational acceleration \cite{PDG2014} and the quantum spatial
scale of UCNs such as $\ell_0 = (2m g^2)^{-1/3}= 5.87\,{\rm \mu m} =
29.75\,{\rm eV^{-1}}$ \cite{Jenke2014,Abele2}. In
Fig.\,\ref{fig:qbounce} we plot the transmission curves of the transitions between
the quantum states shown in Fig.~\ref{fig:qboun}.
\begin{figure}
\centering
\includegraphics[height=0.25\textheight]{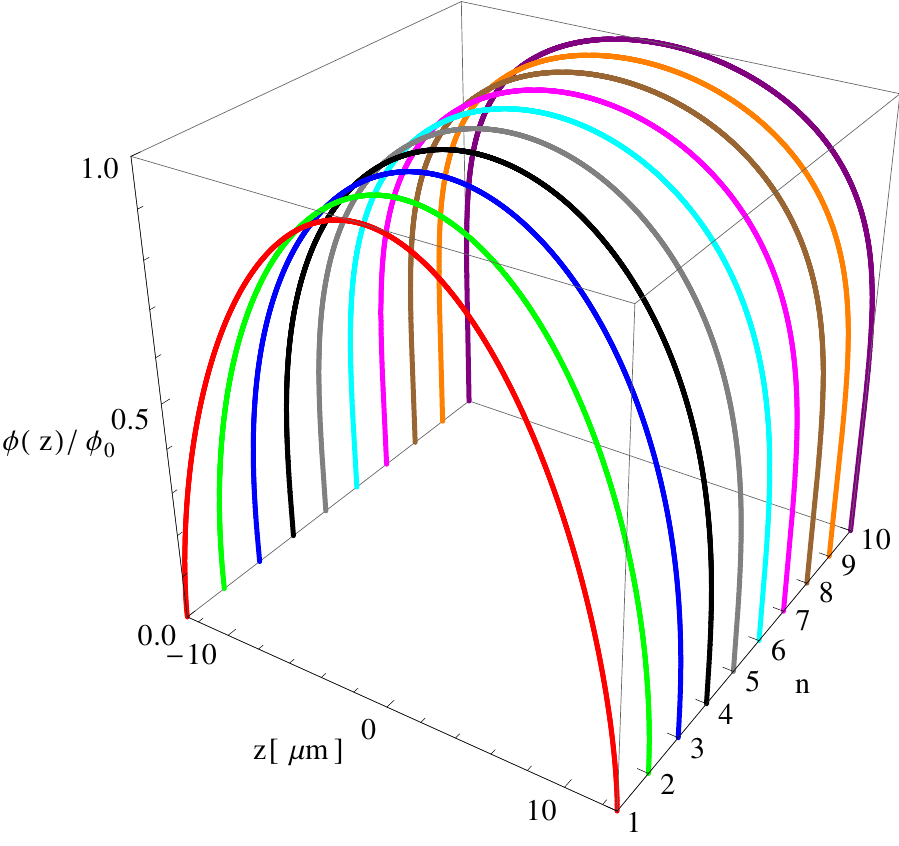}
\caption{The profiles of the chameleon field, calculated in the strong
  coupling limit $\beta > 10^5$ in the spatial region $\textstyle z^2
  \le d^2/4$ for $d = 30.1\,{\rm \mu m}$ and $n \in [1, 10]$ in
  \cite{Ivanov2013} and used for the extraction of the upper bound of
  the coupling constant $\beta$, i.e. $\beta < 5.8 \times 10^8$
  \cite{Jenke2014}.}
\label{fig:profile}
\end{figure}
The extraction of the upper bound of $\beta$, i.e. $\beta < 5.8 \times
10^8$, has been performed within chameleon field theory using the
Ratra--Peebles potential for the chameleon self--interaction
\cite{Chameleon1,Chameleon2,Brax2011,Ivanov2013}. The
profiles of the chameleon field, confined between two mirrors and
separated by a distance $d = 30.1\,{\rm \mu m}$ have been calculated in
\cite{Ivanov2013} and are shown in Fig.\,\ref{fig:profile}.
%\begin{figure}
%\centering
%\includegraphics[height=0.25\textheight]{Stheory.pdf}
%\caption{The dependence of the observation of the potential of the
%  self--interaction of scalar (chameleon) field theory on the
%  sensitivity of the experimental data on the transition frequencies
%  of quantum gravitational states of UCNs, measured in \textit{q}\textsc{Bounce}
%  experiments.}
%\label{fig:shape}
%\end{figure}

\begin{figure}
\floatbox[{\capbeside\thisfloatsetup{capbesideposition={right,bottom},capbesidewidth=5cm}}]{figure}[\FBwidth]
{\caption{The dependence of the observation of the potential of the
  self--interaction of scalar (chameleon) field theory on the
  sensitivity of the experimental data on the transition frequencies
  of quantum gravitational states of UCNs, measured in \textit{q}\textsc{Bounce}
  experiments.}}
{\includegraphics[height=0.3\textheight]{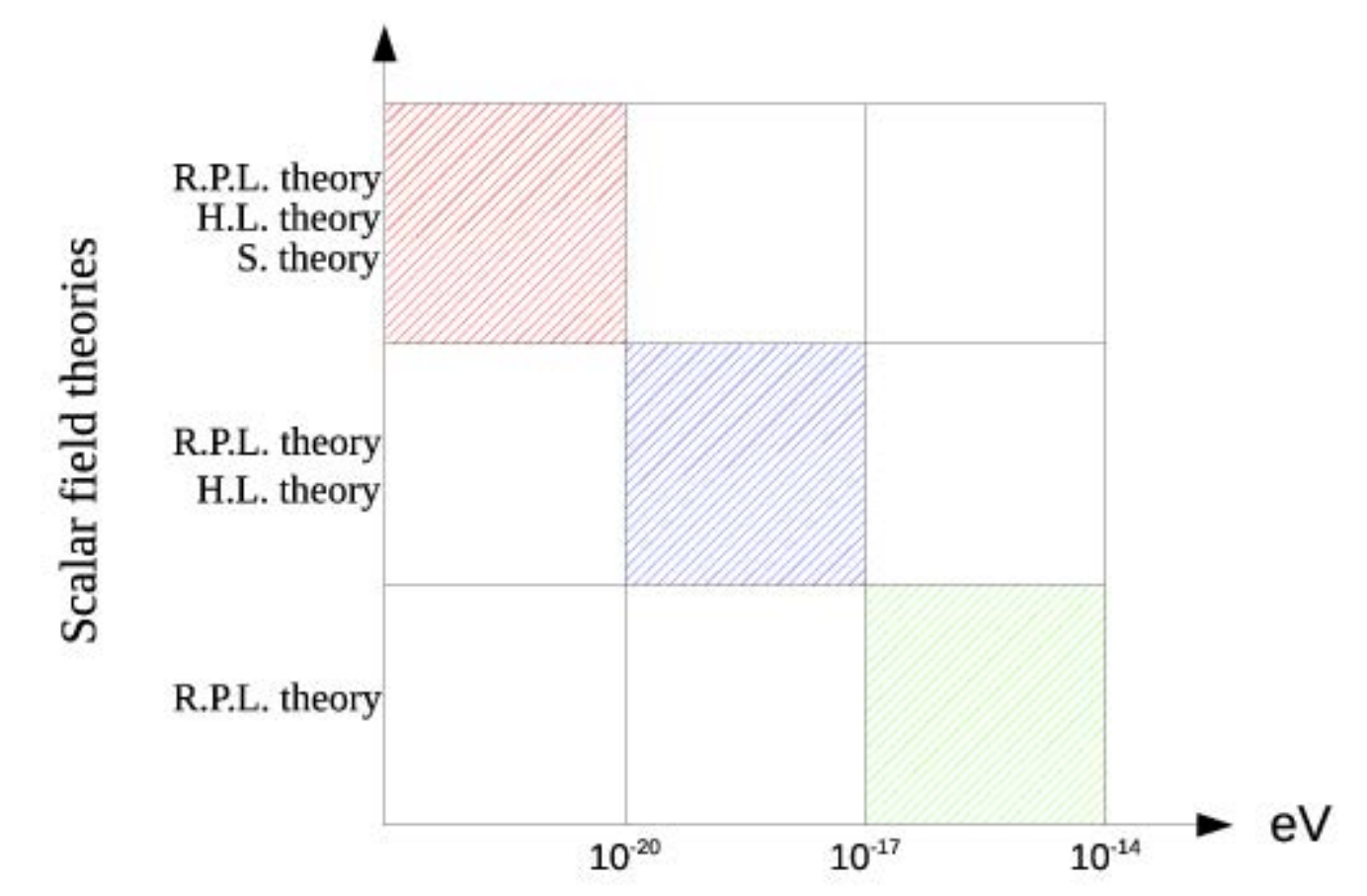}}
\label{fig:shape}
\end{figure}
A precision analysis of the chameleon--matter coupling
constant $\beta$ can be performed by neutron interferometry as proposed by Brax {\it et al.}  \cite{Brax2013,Brax2014} and
has been realized by Lemmel {\it et al.} \cite{Lemmel2015}. Best limits on $\beta$ have been achieved by atom interferometry in~\cite{Ham15}

As it is well known, the Ratra--Peebles potential is just one
possible potential for the self--interaction of scalar fields $\phi$.
The potential can also be taken in the Higgs--like
form \cite{Khoury2006} (see also \cite{Adelberger2007}) and in the
symmetric form \cite{Khoury2010,Upadhye2013}, respectively. The scalar
field with a self--interaction potential, which is symmetric with
respect to a transformation $\phi \to - \phi$, is called {\it
  symmetron}. As it has been shown in \cite{Ivanov2013}, the \textit{q}\textsc{Bounce}
experiments with UCNs are able to distinguish the shape of the
self--interaction potential of the scalar field. In
Fig.\,\ref{fig:shape} we show the dependence of the
shape of the self--interaction potential of the scalar field on the
sensitivity of the experimental data of the \textit{q}\textsc{Bounce} experiments. One
may see that the region of accuracies $\Delta E = (10^{-17} -
10^{-14})\,{\rm eV}$ is sensitive to the Ratra--Peebles potential
only. In turn, the regions of accuracies $\Delta E = (10^{-20} -
10^{-17})\,{\rm eV}$ and $\Delta E < 10^{-20}\,{\rm eV}$ are sensitive
to the scalar field theories with the Higgs--like potential and the
symmetron, respectively.  The sensitivity of about $\Delta E \sim 10^{-21}\,{\rm
  eV}$ is feasible in the \textit{q}\textsc{Bounce} experiments \cite{Abele2}. Hence,
\textit{q}\textsc{Bounce} experiments can be a good tool for measurements of the
effective low--energy torsion--spin--matter (neutron) interactions,
which can be derived from those obtained in \cite{Ivanov2015}. The use
of the \textit{q}\textsc{Bounce} experiments for measurements of torsion--spin--matter
(neutron) interactions should be helpful to overcome the barrier of
extremely small magnitudes of torsion.

The new method profits from small
systematic effects in such systems, mainly due to the fact that in
contrast to atoms, the electric polarisability of the neutron is
extremely low. Neutrons are also not disturbed by short range electric
forces such as van der Waals or Coulomb forces and other
polarisability effects such as the Casimir--Polder interaction of UCNs
with reflecting mirrors. Together with the neutron
neutrality, this provides the key to a sensitivity of several orders
of magnitude below the strength of electromagnetism. A search for a non-vanishing charge of the neutron is also possible.

Hence, experimental measurements of the transition frequencies of
quantum gravitational states of UCNs in the \textit{q}\textsc{Bounce}
experiments \cite{Abele12,Jenke2014,Abele2} and the quantum free fall
of UCNs together with the experimental investigations of the
phase shifts of the wave functions of slow neutrons in neutron
interferometry \cite{Lemmel2015} are very important tools for probing
dark energy and theories of torsion
gravity~\cite{Ivanov2015,Ivanov2014}.

% ****************************************************************************
% BIBLIOGRAPHY AREA
% ****************************************************************************

\begin{footnotesize}

\end{footnotesize}

% ****************************************************************************
% END OF BIBLIOGRAPHY AREA
% ****************************************************************************

\end{document}